\documentclass[3p,times,sort&compress]{elsarticle}

\usepackage{graphicx}
\usepackage{epsfig,amsmath,amssymb,graphicx,pdflscape,color}
\usepackage{hyperref}
\usepackage[noabbrev]{cleveref}
\usepackage{siunitx}
\usepackage{color}
\usepackage[export]{adjustbox}

\crefname{section}{}{}
\crefname{equation}{}{}
\crefname{figure}{}{}
\crefname{table}{}{}
\crefname{appendix}{}{}
\crefname{chapter}{}{}

\newcommand{\vect}[1]{\boldsymbol{#1}}
\DeclareRobustCommand{\e}{\relax\ifmmode\mathrm{e}\else\error\fi}
\usepackage[title]{appendix}
\begin{document}

\begin{frontmatter}
\title{Front stability of infinitely steep travelling waves in population biology}

\author[qut]{Matthew J. Simpson \corref{cor1}}

\author[qut]{Nizhum Rahman}

\author[UniSA]{Alexander K.Y. Tam}

\address[qut]{School of Mathematical Sciences, Queensland University of Technology, Brisbane, Australia.}
\address[UniSA]{UniSA STEM, Mawson Lakes Campus, The University of South Australia, Mawson Lakes, SA 5095, Australia.}
\cortext[cor1]{Corresponding author: matthew.simpson@qut.edu.au}

\begin{abstract}
Reaction--diffusion models are often used to describe biological invasion, where populations of individuals that undergo random motility and proliferation lead to moving fronts.  Many reaction--diffusion models of biological invasion are extensions of the well--known Fisher--KPP model that describes the spatiotemporal evolution of a 1D population density, $u(x,t)$, as a result of linear diffusion with flux  $\mathcal{J} = - \partial u / \partial x$,  and logistic growth source term, $S = u(1-u)$.  In 2020 Fadai introduced a new reaction--diffusion model of biological invasion with a nonlinear degenerate diffusive flux, $\mathcal{J} = - u \partial u / \partial x$, and the model was formulated as a moving boundary problem on $0 < x < L(t)$, with $u(L(t),t)=0$ and $\textrm{d} L(t) / \textrm{d}t = -\kappa u \partial u / \partial x$ at $x = L(t)$ (J Phys A: Math Theor. 53, 095601).  Fadai's model leads to travelling wave solutions with infinitely steep, well--defined fronts at the moving boundary, and the model has the mathematical advantage of being analytically tractable in certain parameter limits.  In this work we consider the stability of the travelling wave solutions presented by Fadai for the first time. We aim to provide general insight by first presenting two key extensions of Fadai's model by considering: (i) generalised nonlinear degenerate diffusion with flux  $\mathcal{J} = - u^m \partial u / \partial x$ for some constant $m > 0$; and, (ii) solutions describing both biological invasion with $\textrm{d} L(t) / \textrm{d}t >0$, and biological recession with $\textrm{d} L(t) / \textrm{d}t<0$.   After establishing the existence of travelling wave solutions for these two extensions, our main contribution is to consider stability of the travelling wave solutions by introducing a lateral perturbation of the travelling wavefront.  Full 2D time--dependent level--set numerical solutions indicate that invasive travelling waves are stable to small amplitude lateral perturbations, whereas receding travelling waves are unstable.  These preliminary numerical observations are corroborated through a linear stability analysis that gives more formal insight into short time growth/decay of wavefront perturbation amplitude.  Julia--based software, including level--set algorithms, is available on \href{https://github.com/alex-tam/OnePhase_PorousFisherStefanStability}{Github} to replicate all results in this study.
\end{abstract}

\end{frontmatter}

\newpage
\section{Introduction}\label{Sec:Intro}
The Fisher--KPP model~\cite{Fisher1937,Kolmogorov1937,Canosa1973}, and generalisations thereof~\cite{Berestycki2017,Murray2002}, are routinely used to understand and interpret the invasion of biological populations.  These continuum partial differential equation (PDE) models typically incorporate a motility mechanism such as linear diffusion, and a source term such as logistic growth to represent a birth-death process~\cite{Fisher1937,Kolmogorov1937,Canosa1973}.  Combined motility and proliferation mechanisms describe the spatiotemporal evolution of a population density $u \ge 0$.  From an applications point of view, these models have been used in many contexts, such as studying populations of trees~\cite{Acevedo2012}, humans~\cite{Steele1998}, rats~\cite{Skellam1951} and biological cells, including modelling normal development and repair~\cite{Sherratt1990,Sengers2007,Maini2004} as well as modelling disease progression~\cite{Swanson2003,Jin2021}.  From a mathematical point of view, applying these models in a 1D Cartesian geometry with appropriate boundary conditions leads to long--time travelling wave solutions~\cite{Fisher1937,Canosa1973,Murray2002,El-Hachem2019}.  Working within the travelling wave coordinate, the long time solution of the time--dependent PDE model can be approximated by the solution of an ordinary differential equation (ODE) that can be studied in the phase plane to provide insight into the original time--dependent PDE model~\cite{Fisher1937,Canosa1973,Murray2002,El-Hachem2019}.

\begin{figure}
\includegraphics[width=1.0\textwidth]{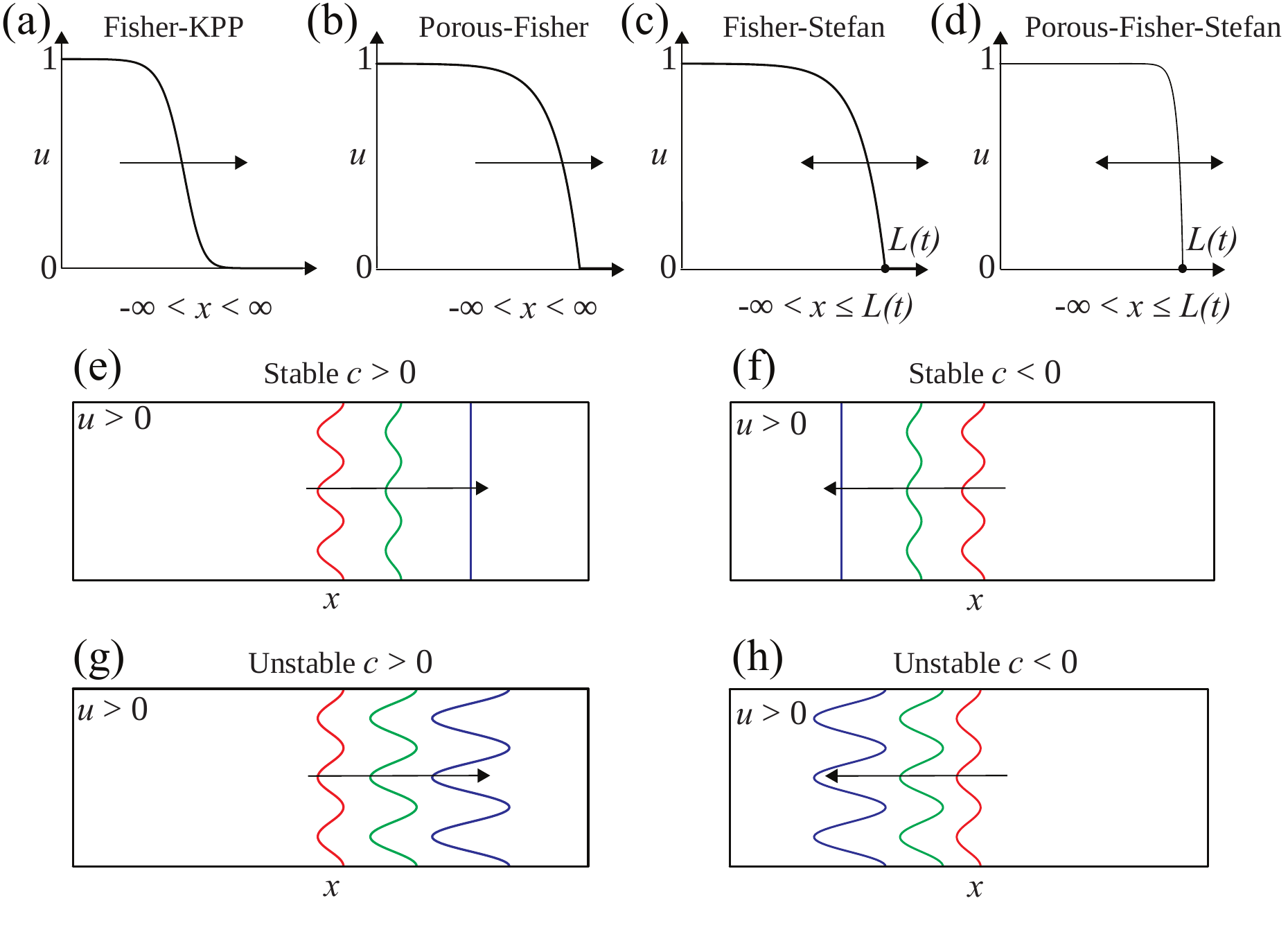}
\caption{Schematic travelling wave solutions: (a) Fisher--KPP model on $-\infty < x < \infty$; (b) Porous--Fisher model on $-\infty < x < \infty$; (c) Fisher--Stefan model on $x < L(t)$ ; and (d) the Porous--Fisher--Stefan (PFS) model on $x < L(t)$. Arrows on the travelling wave profiles in (a)--(d) indicate that the Fisher--KPP and Porous--Fisher models describe biological invasion only, whereas the Fisher--Stefan and PFS models can describe biological invasion and biological recession.  Schematics in (e)--(f) illustrate four possibilities for the linear stability of travelling wave solutions of the PFS model with $u \to 1$ as $x \to -\infty$.  The profile at $t=0$ (red) shows the location of the moving boundary where $u=0$ including a small amplitude transverse perturbation.  Profiles at $t_1 > 0$ (green) and $t_2  > t_1 > 0$ (blue) and shows the evolution of the moving boundary.  Arrows in (e)--(h) show the direction of increasing time.  Profiles in (e)--(f) are stable  invading and receding fronts, respectively.  Profiles in (g)--(h) are  unstable invading and receding fronts, respectively.   \label{Fig1}}
\end{figure}

A schematic of the travelling wave solution of the Fisher--KPP model is shown in Figure \ref{Fig1}(a).  The non-dimensional 1D Fisher--KPP model involves linear diffusion with flux $\mathcal{J} = - \partial u / \partial x$,  and logistic source term $S = u(1-u)$~\cite{Murray2002}.  Long--time travelling wave solutions on $-\infty < x < \infty$ travel with speed $c > 2$, and these solutions do not have compact support since $u > 0$ for all $x$.  This means that these travelling wave solutions cannot be used to model experimental observations that often include a clearly--defined front position~\cite{Maini2004,McCue2019}.  Another limitation of the Fisher--KPP model on $-\infty < x < \infty$ is that any positive initial condition that vanishes as $x \to \pm \infty$ always leads to \textit{invading} travelling waves which means that previously unoccupied regions eventually become occupied.  This means that the Fisher-KPP model on $-\infty < x < \infty$ cannot be used to model biological extinction or biological recession~\cite{El-Hachem2021}.  One approach to deal with the first limitation of the Fisher--KPP model is to generalise the linear diffusion flux to a degenerate nonlinear diffusive flux, $\mathcal{J} = - u \partial u / \partial x$~\cite{Sherratt1990,Maini2004,Sengers2007,Witelski1995,Harris2004,Johnston2023}. This model, sometimes called the \textit{Porous--Fisher} model~\cite{Witelski1995}, leads to sharp--fronted travelling wave solutions on $-\infty < x < \infty$ that move with speed of $c > 1/\sqrt{2}$.  As illustrated in Figure \ref{Fig1}(b), travelling wave solutions of the Porous--Fisher model are sharp--fronted, which means that these solutions can be used to match experimental observations with a clearly--defined front position.  Similar to the Fisher--KPP model, the Porous--Fisher model can only describe invading travelling wave solutions where previously unoccupied regions eventually become occupied~\cite{El-Hachem2021}.

In 2010, Du and Lin~\cite{Du2010} pioneered a different way of overcoming the first limitation of the Fisher--KPP model by re--formulating that model as a moving boundary problem.   This approach, sometimes called the \textit{Fisher--Stefan} model~\cite{El-Hachem2019} involves working with a linear diffusive flux and logistic source term on $-\infty < x < L(t)$ with $u(L(t),t)=0$ and a Stefan--like moving boundary condition, $\textrm{d} L(t) / \textrm{d}t = -\kappa  \partial u / \partial x$ at $x = L(t)$.  The constant $\kappa$ is related to the Stefan number in the classical Stefan problem from the heat and mass transfer literature~\cite{Crank1987,Gupta2017}.  While moving boundary problems are classically associated with melting and solidification processes in the heat and mass transfer literature~\cite{BrosaPlanella2019,Mitchell2010}, there is an increasing awareness and application of these types of moving boundary problems for biological applications~\cite{Gaffney1999,Shuttleworth2019,Dalwadi2020,Jepson2022}.  In the context of biological invasion, the Fisher--Stefan model has the advantage that it leads to sharp--fronted travelling wave solutions, as illustrated in Figure \ref{Fig1}(c), as well as having the flexibility to model both biological invasion with $c > 0$, and biological recession with $c < 0$~\cite{El-Hachem2021}.  Therefore, working with a moving boundary model allows us the flexibility to model processes where some initially occupied regions eventually become vacant.

In 2020, Fadai~\cite{Fadai2020} united the properties of the Porous--Fisher and Fisher--Stefan models by working with a degenerate nonlinear diffusive flux and a logistic source term as a moving boundary problem on $-\infty < x < L(t)$, with $u(L(t),t)=0$ and a Stefan-like moving boundary condition $\textrm{d} L(t) / \textrm{d}t = -\kappa u \partial u / \partial x$ at $x = L(t)$.  This model, called the \textit{Porous-Fisher-Stefan} (PFS) model leads to sharp--fronted travelling waves, as illustrated in Figure \ref{Fig1}(d), with the additional property that the front of the travelling wave is infinitely steep at $x=L(t)$ where $u=0$. Fadai showed, using a combination of phase plane and perturbation methods, that travelling wave solutions of the PFS model are analytically tractable in certain limits, $\kappa \to 0$ and $\kappa \to \infty$, however Fadai did not consider any stability properties of these travelling wave solutions.

In this work we consider the front stability of travelling wave solutions of the PFS model.  Before exploring questions of front stability,  we extend Fadai's travelling wave analysis in two ways by: (i) working with a generalised nonlinear degenerate diffusion model with flux  $\mathcal{J} = - u^m \partial u / \partial x$ for some constant $m > 0$; and (ii) considering solutions describing biological invasion with $\kappa > 0$ and $\textrm{d} L(t) / \textrm{d}t >0$, as well as solutions describing biological recession with $\kappa < 0$ and $\textrm{d} L(t) / \textrm{d}t<0$.   After establishing key properties of travelling wave solutions for this generalised PFS model, we then consider the stability of this broader family of travelling wave solutions by introducing a small amplitude lateral perturbation of the travelling wavefront, as shown schematically in Figure \ref{Fig1}(e)--(h).  In summary, 2D time--dependent level--set numerical solutions indicate that models of invasion with $\kappa > 0$ are stable to small lateral perturbations, whereas  models of biological recession with $\kappa < 0$ are unstable.  These preliminary numerical observations are corroborated and formalised using linear stability analysis that describes short time growth or decay of the perturbation amplitude.  Julia--based software, including the level-set algorithms, is available on \href{https://github.com/alex-tam/OnePhase_PorousFisherStefanStability}{Github} to replicate all results in this study.

\newpage

\section{Mathematical Model}
\subsection{Two--dimensional Porous-Fisher-Stefan model}\label{Sec:PDE}
To explore stability of travelling wave solutions of the PFS model we first re--cast the reaction--diffusion equation on a 2D domain $\Omega (t)$, with a moving boundary, $\partial \Omega (t)$.  Within this context the non-dimensional 2D PFS model can be written as
\begin{subequations}
\label{eq_main_2d}
\begin{align}
    &\frac{\partial u}{\partial t}= \frac{\partial}{\partial x}\left[u^m \frac{\partial u}{\partial x}\right]+ \frac{\partial}{\partial y}\left[u^m \frac{\partial u}{\partial y}\right]+ u(1-u) \;\; \text{on}   \;\;   \Omega (t),\label{Eq:PFS}\;\\
     &u(x,y,t)=0 \;\; \text{on}   \;\;  \partial \Omega (t),  \\
     &u(x,y,0)=U(x,y) \;\; \text{on}   \;\; \Omega (0), \label{Eq:PFS IC}\\
&\mathcal{V}=-\kappa u^m \nabla u \cdot \hat{n}\;\;  \text{on}   \;\;  \partial \Omega (t), \label{Eq:PFS MovingBC}
\end{align}
\end{subequations}
where $U$ is the initial population density, $\mathcal{V}$ is the normal velocity of the interface, $\kappa$ is a constant that determines the speed of the moving boundary, and $m > 0$ is the exponent associated with the power law nonlinear diffusivity.  Note that Equations (\ref{Eq:PFS})--(\ref{Eq:PFS MovingBC}) are identical to the mathematical model studied by Fadai~\cite{Fadai2020} when $m=1$ and $u(x,y,t)$ depends only on $x$ and $t$.   Therefore, we view Equations (\ref{Eq:PFS})--(\ref{Eq:PFS MovingBC}) as a relatively straightforward extension of Fadai's model.  The main implication of the generalisations introduced here is that the boundary fixing transformation used by Fadai to solve the 1D time--dependent PDE model is not applicable in 2D. Later in Section \ref{Sec:NumericalMethod} we will describe how to solve this 2D time--dependent PDE model using a level--set method~\cite{Sethian1999,Osher2003,Tam2023}.

Note that the moving boundary condition, Equation (\ref{Eq:PFS MovingBC}), indicates that the front velocity is proportional to $u^m \nabla u$, and we have $u=0$ on the moving boundary, $\partial \Omega (t)$.  Therefore, if $\mathcal{V}$ is finite, the gradient of $u$ at the moving boundary will be infinite, as illustrated schematically in Figure \ref{Fig1}(d) and established mathematically by Fadai~\cite{Fadai2020} for $m=1$.  To deal with this complication numerically we introduce a transformation $\phi=u^{m+1}$ to give
\begin{subequations}
  \label{eq:theta_main}
\begin{align}
    \frac{\partial \phi}{\partial t}&=  \phi^{m/(1+m)} \left(\frac{\partial^2 \phi}{\partial x^2}+ \frac{\partial^2 \phi}{\partial y^2} \right) +(1+m) \phi \left(1-\phi^{1/(1+m)}\right)\label{Eq:PFSphi}\; \;\; \text{on}   \;\;   \Omega (t),\\
    &\phi(x,y,t)=0 \;\; \text{on}   \;\;  \partial \Omega (t), \\
    &\phi(x,y,0)=U(x,y)^{m+1} \;\; \text{on}   \;\; \Omega (0),  \label{Eq:PFSphi IC}\\
&\mathcal{V}=-\frac{\kappa}{1+m}\nabla \phi \cdot \hat{n}\;\;  \text{on}   \;\;  \partial \Omega (t), \label{Eq:PFSphi MovingBC}
\end{align}
\end{subequations}
which avoids the indeterminate form of Equation (\ref{Eq:PFS MovingBC}).  Therefore, we will solve Equations (\ref{Eq:PFS})--(\ref{Eq:PFS MovingBC}) by working in the transformed $\phi$ variable and solving Equations (\ref{Eq:PFSphi})--(\ref{Eq:PFSphi MovingBC}) before applying the inverse transform $u = \phi^{1/(m+1)}$ to give $u$. We now present details of a level--set numerical method to achieve this.

\subsection{Level set method}\label{Sec:NumericalMethod}
We solve Equations (\ref{Eq:PFSphi})--(\ref{Eq:PFSphi MovingBC}) on a 2D rectangular computational domain $\mathcal{D}$ using a level--set method~\cite{Sethian1999,Osher2003,Tam2023}. The interface $\partial \Omega(t)$ is embedded as the zero level-set of a scalar function $\varphi(x,y,t)$.  The interface implicitly defined by $\partial \Omega(t)=\{(x,y)\mid \varphi(x,y,t)=0 \}$, where $\varphi(x,y,t)$ is defined everywhere in $\mathcal{D}$ with the key property that $\varphi<0$ on $\Omega(t)$.  To ensure that $\varphi=0$ is maintained on the interface, $\varphi$ must satisfy the following evolution equation
\begin{equation}\label{eq:Velocity_field}
    \frac{\partial \varphi}{\partial t}+F\mid\nabla\varphi\mid=0,
\end{equation}
where $F(x,y,t)$ is the extension velocity field that is defined across $\mathcal{D}$, and satisfies $F=\mathcal{V}$ on  $\partial \Omega(t)$. There are several options for choosing $F$.  Here we define $F$ by estimating $\mathcal{V}$ at the interface using a second-order finite difference approximation and then use orthogonal extrapolation to define $F$ across $\mathcal{D}$~\cite{Osher2003,Tam2023}.

With this level-set function we rewrite Equations (\ref{Eq:PFSphi})--(\ref{Eq:PFSphi MovingBC}) as
\begin{subequations}
 \label{eq:level_system}
\begin{align}
    \frac{\partial \phi}{\partial t}&= \phi^{m/(1+m)} \left(\frac{\partial^2 \phi}{\partial x^2}+ \frac{\partial^2 \phi}{\partial y^2}\right)+(1+m) \phi \left(1-\phi^{1/(1+m)}\right )\label{Eq:PFSLevelSet}\; \; \text{on}   \;\;   \varphi(x,y,t)<0,\\
    &\frac{\partial \varphi}{\partial t}+F \mid \nabla \varphi \mid \, = 0\;\;\text{on}   \;\;   (x,y)\in\mathcal{D}, \label{Eq:PFSLevelSetVelocity}\\
    &\phi(x,y,t)=0 \;\; \text{on}   \;\;  \varphi(x,y,t)=0, \\
    &\phi(x,y,0)=U(x,y)^{m+1} \;\; \text{on}   \;\; \varphi(x,y,0)<0,\\
    &F=-\frac{\kappa}{1+m}\nabla \phi \cdot \hat{n}\;\;  \text{on}   \;\;  \varphi(x,y,t)=0. \label{Eq:PFSLevelSet_Extension}
\end{align}
\end{subequations}

Note that we have attempted to use meaningful and appropriate notation in this Section.  In the level--set literature it is standard to use the variable $\phi$ to denote the level set function~\cite{Sethian1999}, however in this work we have used $\varphi$ for the level--set function.  We make this choice because we use $\phi$ to denote the transformed variable $\phi=u^{m+1}$ to be consistent with  Fadai~\cite{Fadai2020}.

We solve Equations (\ref{Eq:PFSLevelSet})--(\ref{Eq:PFSLevelSet_Extension}) by discretising $\mathcal{D}$ using a standard uniformly spaced finite difference mesh.  Terms on the right-hand side of Equation (\ref{Eq:PFSLevelSet}) are discretised using second-order finite-difference approximation using non-constant grid spacing. Interpolation is used when $\partial \Omega(t)$ lies between mesh points~\cite{Tam2022a}.

The extension velocity field $F$ is determined by approximating Equation (\ref{Eq:PFSLevelSet_Extension}) using a second-order finite difference approximation and orthogonal extrapolation~\cite{Osher2003,Aslam2004}.  Given our estimate of $F$  we approximate Equation (\ref{Eq:PFSLevelSetVelocity}) using a central scheme \cite{Jiang1998,Simpson2005}. With these approximation in place we then solve the resulting system of coupled ODEs numerically using the DifferentialEquations package in Julia~\cite{Tsitouras2011,Rackauckas2017}.

\newpage
\section{Results and discussion} \subsection{Travelling wave solutions: Numerical explorations}
\begin{figure}
\includegraphics[width=1.0\textwidth]{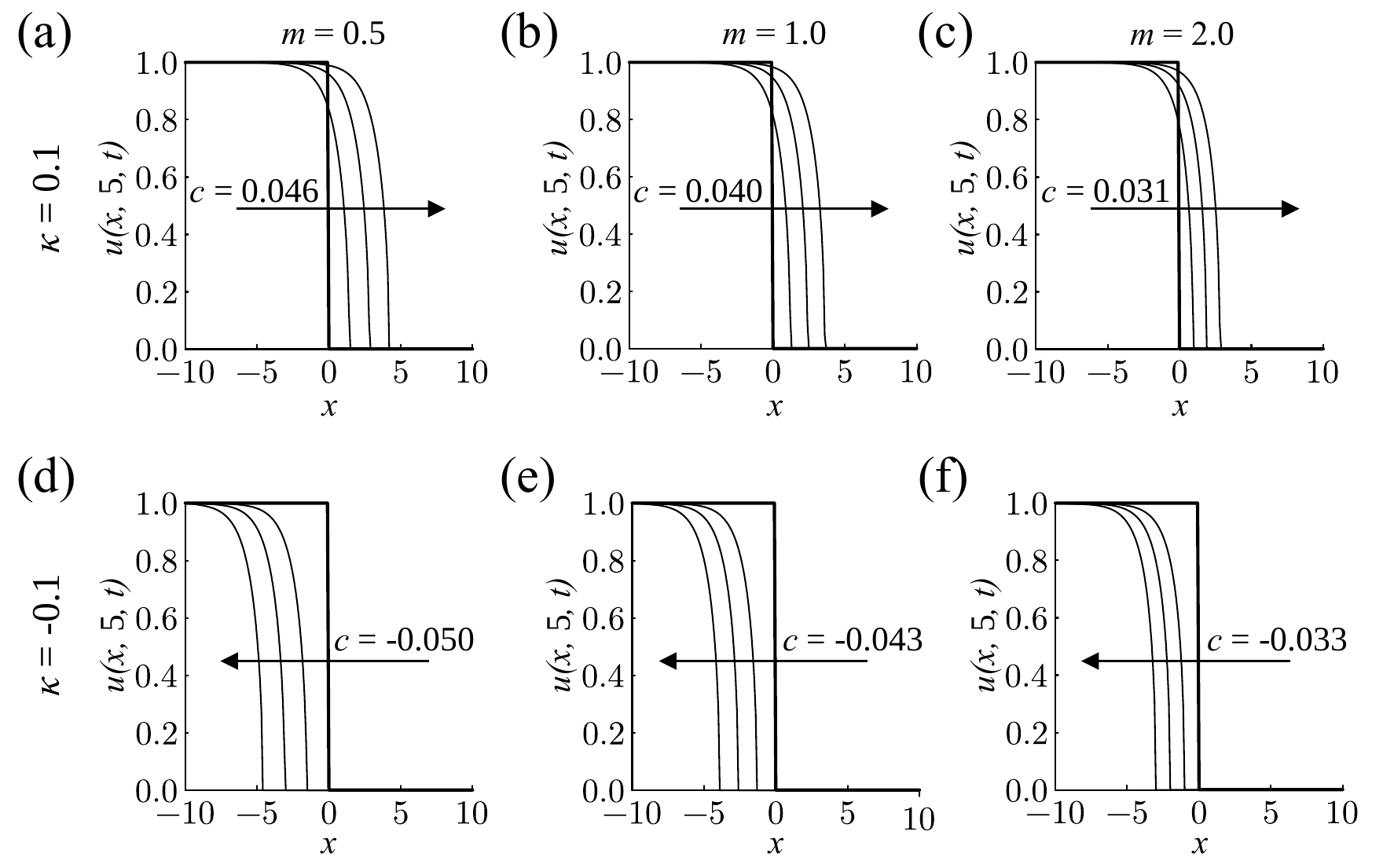}
\caption{Numerical solution of Equations (\ref{Eq:PFS})--(\ref{Eq:PFS MovingBC} on a rectangular domain with $-10 < x < 10$ and $0 < y < 10$, and  $u(x,y,0) = 1$ for $x < 0$ with $\partial \Omega(0)$ along the vertical line at $x=0$.  Numerical simulations are performed with periodic boundary conditions along the horizontal boundaries at $y=0$ and $y=10$, and we set $u=1$ along the vertical boundary at $x=-10$.  Numerical solutions are obtained on a $401 \times 201$ uniform finite difference mesh.  In each plot we show solutions at $t=0, 30, 60$ and $90$ with the arrow showing the direction of increasing time.  Results in (a)--(c) correspond to $\kappa=0.1$ whereas results in (d)--(f) correspond to $\kappa = -0.1$. The value of the exponent $m$ is indicated on each column.  We report the value of the front velocity at the end of the simulation, $c$.  \label{Fig2}}
\end{figure}

We begin by exploring numerical solutions of Equations (\ref{Eq:PFS})--(\ref{Eq:PFS MovingBC}) to generate preliminary numerical evidence of travelling wave solutions for a range of $m$ and $\kappa$ values that have not been considered previously.   Numerical solutions are generated on a rectangular domain with $-10 < x < 10$ and $0 < y < 10$.  The initial condition is $u(x,y,0) = 1$ for $x < 0$, and $\partial \Omega(0)$ is chosen to coincide with the vertical line at $x=0$.   Numerical simulations are performed with periodic boundary conditions at $y=0$ and $y=10$, and we set $u=1$ along the vertical boundary at $x=-10$.  The combination of periodic boundary conditions at $y=0$ and $y=10$ and specifying the initial condition $u(x,y,0)$ to be independent of $y$ means that these 2D simulations are independent of the vertical position and our numerical solutions for $u(x,y,t)$ is also independent of vertical position~\cite{Simpson2023}.  Results are plotted in Figure \ref{Fig2} showing the time evolution of $u(x,5,t)$ for a range of $m$ and $\kappa$.

Results in Figure \ref{Fig2}(a)--(c) show the evolution of the front position for $\kappa > 0$ for $m=0.5, 1.0, 2.0$, respectively.  In each case we see that the initial step function evolves into a front that moves in the positive $x$-direction.  The moving front appears to become approximately constant shape and constant speed over the interval shown, $0 \le t \le 90$.  The level--set method tracks the position and velocity of $\partial \Omega(t)$ and we report the front velocity at $t=90$ as an approximation of the long--time travelling wave speed $c$.   Here we see that setting $\kappa > 0$ leads to an invasion front with $c > 0$, and we see that $c$ decrease with $m$.  Results in Figure \ref{Fig2}(d)--(f) show the front evolution for $\kappa < 0$ for $m=0.5, 1.0, 2.0$, respectively.  In these cases the initial step function evolves into a front that moves in the negative $x$-direction with $c <0$~\cite{El-Hachem2021}, and again our numerical estimates of $c$ increases with $m$.

The numerical exploration in Figure \ref{Fig2} extends the kinds of travelling waves considered previously by Fadai who focused on $m=1$ and $\kappa > 0$~\cite{Fadai2020}.  Here we see that different values of $m$ also lead to travelling wave solutions for the PFS model, and that setting $\kappa < 0$ leads to receding travelling waves.  We now turn to the phase plane to provide mathematical insight into these travelling wave solutions, as well as enabling us to gather additional quantitative information to check the accuracy of our level--set PDE code.   It is notable that our estimates of $c$ from the time--dependent PDE model involves specifying $\kappa$ and $m$ as inputs into numerical solution, and we then use the late--time front velocity to provide an estimate of $c$ as an output.  This calculation is approximate in the sense that we necessarily introduce truncation error by discretizing $\mathcal{D}$, and then estimate the long--time travelling wave speed $c$ using a finite--time simulation.  In Section \ref{Sec:PP} we will use the phase plane provide a more formal argument about the existence of travelling wave solutions, as well as using the associated dynamical system to provide an accurate and independent unique value of $c$ for each combination of $\kappa$ and $m$.

\subsection{Phase plane analysis}\label{Sec:PP}
To analyze the travelling wave solutions depicted in Figure \ref{Fig2} we consider the 1D time-dependent PDE model for $\phi(x,t)$,
\begin{equation}
    \dfrac{\partial \phi}{\partial t}=  \phi^{m/(1+m)} \dfrac{\partial^2 \phi}{\partial x^2} +(1+m) \phi \left(1-\phi^{1/(1+m)}\right)\label{Eq:PFSphi1D}
\end{equation}
on $-\infty < x < L(t)$, with $\phi(L(t),t)=0$ and $\displaystyle{\lim_{x \to -\infty}\phi(x,t) =1}$.  In the 1D coordinate system the evolution of the moving boundary is given by $\displaystyle{\mathrm{d}L(t)/ \textrm{d}t= - \kappa \partial \phi/ \partial x /(m+1)}$ evaluated  at $x = L(t)$.  We now seek travelling wave solutions in the usual way by re-writing $\phi(x,t) = \phi(z)$, where $z = x -ct$, and noting that $\displaystyle{\textrm{d}L/\textrm{d}t=c}$ in the long time limit giving
\begin{equation}
\phi(z)^{m/(1+m)}\;\phi''(z) +c\;\phi'(z)+ (1+m)\;\phi(z)\;\left(1-\phi(z)^{1/(1+m)} \right)=0 \label{Eq:TravellingWave}
\end{equation}
on $-\infty < z < 0$.  Here, derivatives with respect to $z$ are written with the prime notation, and the boundary conditions are $\displaystyle{\lim_{z \to -\infty}\phi(z) =1}$, $\phi(0)=0$ and $c(1+m)= -\kappa \phi'(0)$.

To proceed we write Equation (\ref{Eq:TravellingWave}) as a system of two first order ODEs
\begin{align}
&\phi'(z)=\psi(z),\label{eq:theta_prime}\\
&\psi'(z)= -\frac{c\;\psi(z)}{\phi^{m/(1+m)}}- (1+m)\;\phi^{1/(1+m)}\left( 1-\phi^{1/(1+m)}\right),\label{eq:psi_prime}
\end{align}
which is identical to the system considered by Fadai~\cite{Fadai2020} when $m=1$.  This system is characterised by two equilibrium points: $(0,0)$ and $(1,0)$.  It is interesting to note that standard phase plane analysis of travelling wave solutions of  the Fisher-KPP model involves establishing the existence of a heteroclinic orbit between two equilibrium points~\cite{Murray2002}, however the phase plane analysis for travelling wave solutions of the PFS model is quite different as travelling wave solutions do not correspond to a heteroclinic orbit in the phase plane.  Instead, travelling wave solutions of the PFS model correspond to a trajectory that leaves $(1,0)$ and moves through the fourth quadrant of the phase plane to intersect the $\psi$ axis at some point $\psi^* < 0$  which corresponds to the moving boundary condition. Linearisation confirms that $(1,0)$ is a saddle point for all $c$ and $m$, and there is always a trajectory leaving $(1,0)$ that moves along the unstable manifold into the fourth quadrant of the $(\phi,\psi)$ phase plane.

\begin{figure}
\includegraphics[width=1.0\textwidth]{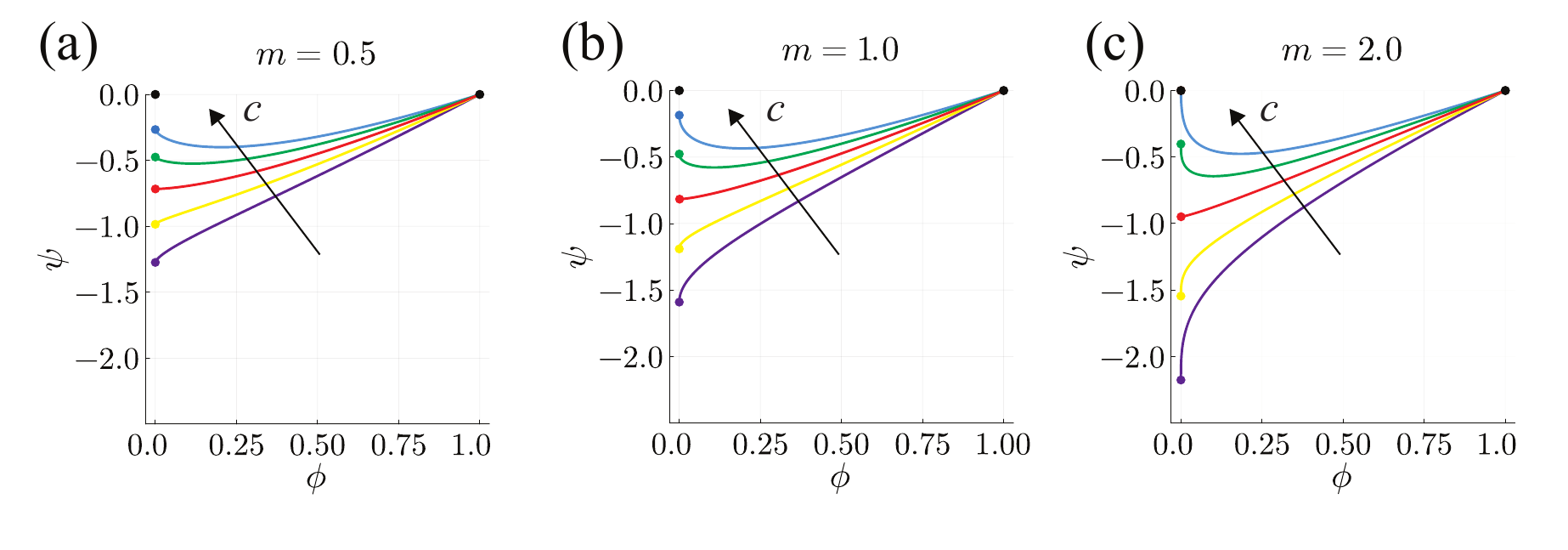}
\caption{\label{Fig3} Phase planes and associated trajectories for Equations (\ref{eq:theta_prime})--(\ref{eq:psi_prime}).  Phase planes in (a)--(c) correspond to $m=0.5, 1.0, 2.0$ as indicated, and each phase plane shows the location of the equilibrium points at $(0,0)$ and $(1,0)$ with black discs.  Each phase plane shows trajectories associated with travelling wave solutions for $c=-0.5, -0.25, 0, 0.25, 0.5$ in purple, yellow, red, green and blue, respectively.  Each trajectory intersects the $\psi$ axis at a particular point shown in a coloured dot.  The coordinates of this point allow us to calculate $\kappa$ according to $\kappa = -c(1+m)/ \psi(0)$. For $c=-0.5, -0.25, 0, 0.25, 0.5$ and $m=0.5$ we obtain $\kappa=-0.59,-0.38,0.0,0.79, 2.82$, respectively. For $c=-0.5, -0.25, 0, 0.25, 0.5$ and $m=1.0$ we obtain $\kappa=-0.63,-0.42,0.0,1.05, 5.4$, respectively.  For $c=-0.5, -0.25, 0, 0.25, 0.5$ and $m=2.0$ we obtain $\kappa=-0.69,-0.49,0.0,1.87, 290241.59$, respectively.}
\end{figure}

Results in Figure \ref{Fig3} visualise several phase planes and associated trajectories for various values of $c$ and $m$. Each phase plane shows the two equilibrium points at $(0,0)$ and $(1,0)$, and a series of trajectories obtained by solving Equations (\ref{eq:theta_prime})--(\ref{eq:psi_prime}) numerically using the DifferentialEquations package in Julia~\cite{Rackauckas2017}.  The initial point on each trajectory is chosen to be arbitrarily close to $(1,0)$ on the unstable manifold in the fourth quadrant.  The associated phase plane trajectory, $(\phi(z),\psi(z))$, is obtained by solving Equations (\ref{eq:theta_prime})--(\ref{eq:psi_prime}) numerically until each trajectory crosses the $\psi$ axis.  The point at which each trajectory crosses the $\psi$ axis, $(0,\psi^*)$ is highlighted for each trajectory, and the coordinates of this point allow us to calculate $\kappa$ by rearranging the moving boundary condition to give $\kappa = -c(1+m)/ \psi^*$.  Therefore, when we use the phase plane we consider $c$ and $m$ and inputs and $\kappa$ as an output.

Phase plane results in Figure \ref{Fig3} allow us to test the accuracy of our time--dependent PDE solutions in Figure \ref{Fig2} in the following way.  In Figure \ref{Fig2} we report several estimates of the travelling wave speed $c$ as the long--time outcome of a time--dependent PDE solution for six choices of $\kappa$ and $m$.  For example, Figure \ref{Fig2}(a)--(c) we estimate $c = 0.046, 0.040, 0.031$ when $\kappa=0.1$ for $m=0.5, 1.0, 2.0$, respectively.    Generating phase planes and associated trajectories for $(c,m)=(0.046, 0.5), (0.040, 1.0), (0.031, 2.0)$ gives $\kappa =0.1029, 0.1053, 0.1059$, respectively. These three values of $\kappa$ are relatively close to $\kappa=0.1$ that was used to generate time--dependent PDE solutions using the level--set method.  This comparison is useful because it provides an independent check on the accuracy of our level-set method.  In a similar way, results in Figure \ref{Fig2}(d)--(f) provide estimates of $c = -0.050, -0.043, -0.033$ when $\kappa=-0.1$ for $m=0.5, 1.0, 2.0$, respectively.    Generating phase planes and associated trajectories for $(c,m)=(-0.050, 0.5), (-0.043, 1.0), (-0.033, 2.0)$ leads to $\kappa =-0.9754, -0.9788, -0.9658$, respectively.  Again, these estimates are consistent with out time--dependent PDE solutions.  We anticipate that the match between the time--dependent PDE solutions and the phase plane results would improve if we refine the mesh used to solve the PDE model and truncation effects are reduced.

So far, we have interpreted the trajectories in Figure \ref{Fig3} as parametric curves,  $(\phi(z),\psi(z))$.  An alternative approach is to re-write Equations (\ref{eq:theta_prime})--(\ref{eq:psi_prime}) using the chain rule to give
\begin{align}
    \frac{\mathrm{d}\psi}{\mathrm{d}\phi}&=-\dfrac{c}{\phi^{{m}/({1+m})}}- \dfrac{(1+m)\;\phi^{{1}/({1+m})}\left( 1-\phi^{{1}/({1+m})}\right)}{\psi(\phi)},\label{eq:trajectories}
\end{align}
whose solution, $\psi(\phi)$, directly describes the shape of the trajectories.  The relevant solution of Equation (\ref{eq:trajectories}) must satisfy $\psi(1)=0$, and then the value of $\kappa$ can be determined by applying $\psi(0) = -c(1+m)/\kappa$.  Although we have been unable to  find any exact solution for the general case of Equation (\ref{eq:trajectories}), we are able to make progress analytically for both slowly invading and slowly receding travelling waves. For $|c| \ll 1$, we expand $\psi(\phi)$ as a regular perturbation expansion in $c$, giving $\psi(\phi)=V_0(\phi)+cV_1(\phi)+\mathcal{O}(c^2)$. Substituting this expansion into \eqref{eq:trajectories} gives:
\begin{align}
    &\mathcal{O}(1):\quad\; -V_0(\phi)\;\frac{\mathrm{d}V_0(\phi)}{\mathrm{d}\phi}=(1+m)\phi^{{m}/({1+m})}\left(1-\phi^{{m}/({1+m})}\right), \quad V_0(1)=0, \label{eq:order_1}\\
    &\mathcal{O}(c):\quad\; -V_0(\phi)\;\frac{\mathrm{d}V_1(\phi)}{\mathrm{d}\phi}-V_1(\phi)\;\frac{\mathrm{d}V_0(\phi)}{\mathrm{d}\phi}= \frac{V_0(\phi)}{\phi^{{m}/({1+m})}}, \quad V_1(1)=0.\label{eq:order_c}
\end{align}
The solution of Equations (\ref{eq:order_1})--(\ref{eq:order_c}) are
\begin{align}
    V_0^2(\phi)&=\dfrac{2(1+m)^2}{6m^2 + 5m + 1}\left[(3m + 1)\phi^{({2m+1})/({1+m})} - (2m + 1)\phi^{({3m+1})/({1+m})} - m\right], \\
    V_1(\phi)&=-\dfrac{1}{V_0(\phi)}\int_{\phi}^{1} V_0(s)\;s^{-{m}/({1+m})} \,\mathrm{d}s.
\end{align}
While we have obtained a closed form expressions for $V_0(\phi)$ for arbitrary $m$, the integral in Equation (\ref{eq:order_c}) appears to have a closed--form expression for $m=1$ only. For the special case where $m=1$ we can evaluate these expressions to give a two-term perturbation approximation, $\psi(\phi)=V_0(\phi)+cV_1(\phi)+\mathcal{O}(c^2)$.  Substituting $\phi=0$ into this approximate solution gives an expression for $c(\kappa)$, or equivalently $\kappa(c)$,
\begin{align}
\label{eq:c_kappa_perturbed_relation}
    c(\kappa) =  \frac{27 \kappa \sqrt{2}}{54\sqrt{3}+\alpha \kappa},\quad  \kappa(c) = \frac{54c\sqrt{3} }{27\sqrt{2}-\alpha c},
\end{align}
where $\alpha=36 \sqrt{2}-6 \sqrt{3}+24\; \mathrm{log}[(\sqrt{3}-1)/(3\sqrt{2}-4)]$~\cite{Fadai2020}.

\begin{figure}
\begin{center}
\includegraphics[width=0.8\textwidth]{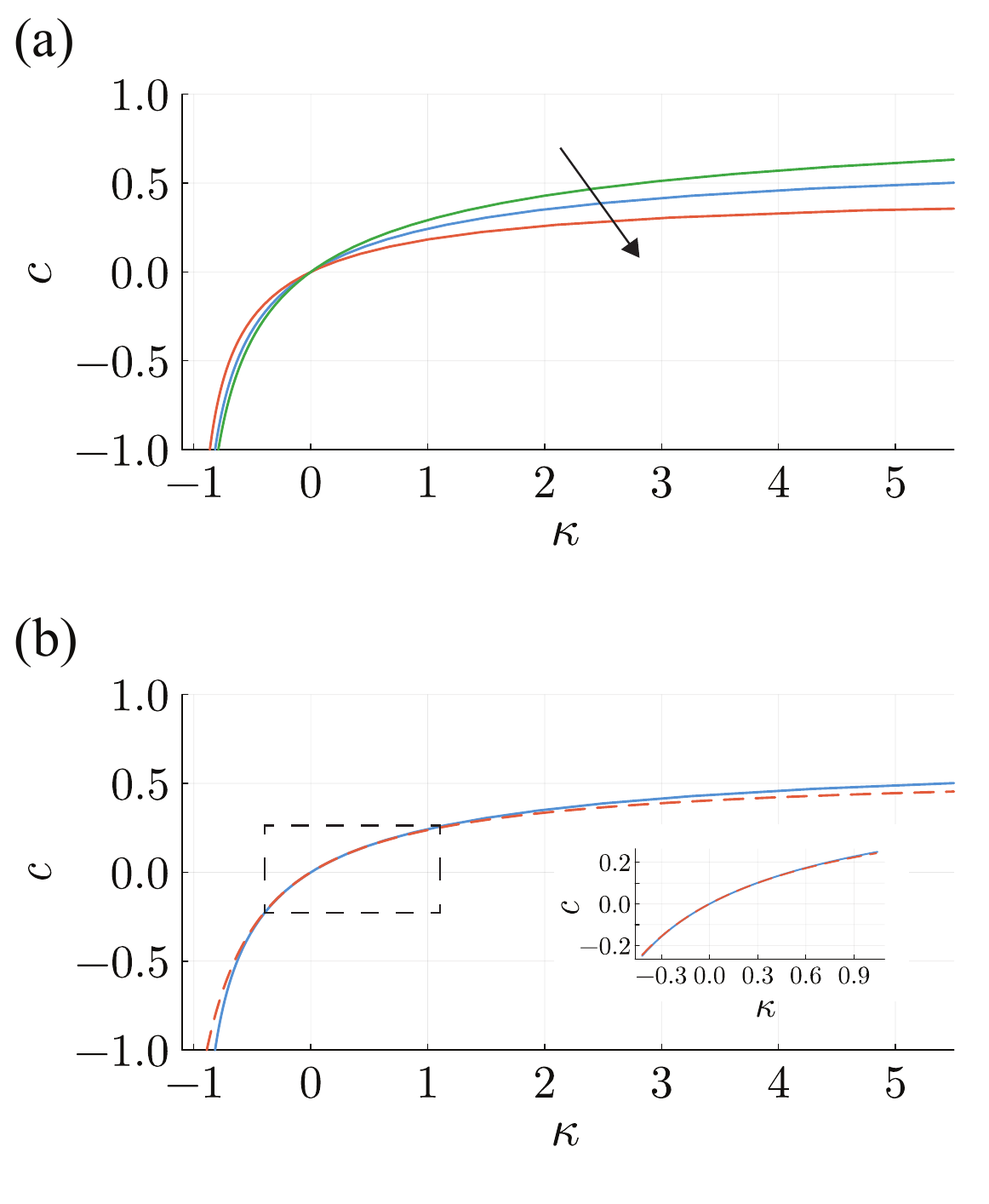}
\caption{\label{Fig4}Relationships between $c$, $\kappa$ and $m$ deriving using the phase plane.  Results in (a) show $c$ as a function of $\kappa$ for $m=0.5, 1.0, 2.0$ in green, blue and red curves, respectively, and the arrow shows the direction of increasing $m$.  Results in (a) are obtained by solving Equations (\ref{eq:theta_prime})--(\ref{eq:psi_prime}) for certain values of $c$ and $m$, and then applying the boundary condition $\kappa =-c(1+m)/ \psi^*$ to estimate $\kappa$.  Results in (b) are for the special case $m=1$ where the solid blue curve is identical to the solid blue curve in (a), and the dashed red curve is the perturbation result, Equation (\ref{eq:c_kappa_perturbed_relation}).  The inset shows a plot that focuses on the dashed rectangular region near $c=0$ where we see that the perturbation result is visually indistinguishable from the numerically--generated phase plane results at this scale.  Each curve in (a)--(b) obtained using the phase plane is constructed by using 50 equally--spaced values of $c$ for each $m$, and then using the phase plane to solve for $\kappa$.   Interpolating the $(c, \kappa)$ values gives the continuous curves in (a)--(b).}
\end{center}
\end{figure}

These various phase plane tools and results allow us to investigate the relationship between $\kappa$, $m$ and $c$.  Our investigation of this relationship is relevant for both invading travelling waves with $c>0$ and receding travelling waves with $c < 0$.  Firstly, and more generally, we can solve Equations (\ref{eq:theta_prime})--(\ref{eq:psi_prime}) numerically to provide various phase plane trajectories that allows us to relate $c$, $\kappa$ and $m$ as we have shown in Figure \ref{Fig4}(a) where we see that $c$ is an increasing function of $\kappa$ for various $m$.  These curves recover the obvious result that we have $c=0$ for $\kappa=0$, regardless of $m$.  Together these results show that invading travelling wave solutions are associated with $\kappa > 0$ whereas receding travelling wave solutions are associated with $\kappa < 0$.

For the special case of $m=1$ we can also generate various phase plane trajectories to relate $c$ and $\kappa$ in the same way that we did for the results in Figure \ref{Fig4}(a), allowing us to relate $c$ and $\kappa$ for any value of $\kappa$.   In addition, for $m=1$ we also have expressions for $c(\kappa)$ (or $\kappa(c)$) relevant for  $|c| \ll 1$ given by Equation (\ref{eq:c_kappa_perturbed_relation}).  Results in Figure \ref{Fig4}(b) compare our perturbation approximation with the more general relationship from the phase plane where we see that the two curves are visually indistinguishable near $c=0$.  Again, it is interesting to note that  Equation (\ref{eq:c_kappa_perturbed_relation}) holds for both slowly invading $c>0$ and slowly receding $c<0$ travelling waves.

\subsection{Travelling wavefront stability: Numerical explorations} \label{Sec:Stability_Numerical}
To address the question of whether the travelling wave solutions explored in Sections \ref{Sec:PDE}--\ref{Sec:PP} are stable to transverse perturbations we first consider generating some illustrative time--dependent PDE solutions.  We initialise these numerical solutions by first solving the boundary value problem that defines the shape of the usual travelling wave $u(z)$ for various choices of $\kappa$ and $m$ using a numerical approach outlined in Section \ref{Sec:Stability_LSA}.  These profiles are used to initiate several time--dependent PDE solutions on a square domain with $0 \le x \le 10$ and $0 \le y \le 10$.  The travelling wave initial condition is shifted so that the vertical moving boundary is located half way along the horizontal extent of the domain at $x=5$, and we then introduce a small transverse sinusoidal perturbation with wavenumber $q$ to give the initial condition we use to solve  Equations (\ref{Eq:PFSphi})--(\ref{Eq:PFSphi MovingBC}).  For these numerical simulations we specify periodic boundary conditions along $y=0$ and $y=10$, and we fox $u=1$ along $x=0$.

\begin{figure}
\includegraphics[width=1.0\textwidth]{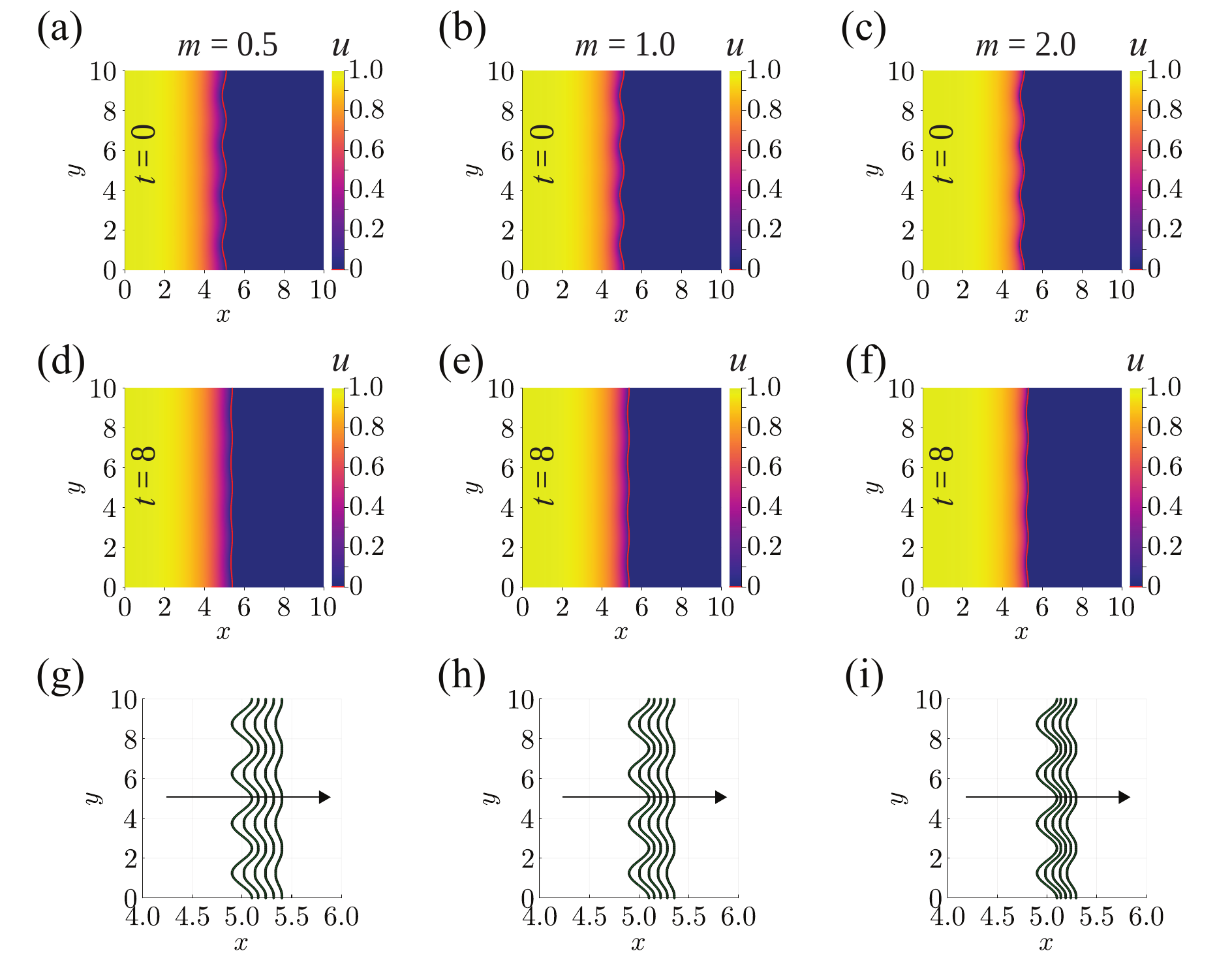}
\caption{\label{Fig5} Numerical exploration of front stability with $\kappa = 0.1$.  Profiles in (a)--(c) show three perturbed travelling wave profiles for $m=0.5, 1.0, 2.0$, respectively.  Each perturbed travelling wave profile is computed by first solving the $\mathcal{O}(1)$ boundary value problem to give the shape of the unperturbed travelling wave that is then shifted so that $\partial \Omega(0)$ coincides with the vertical line at $x=5$ before a perturbation with wavenumber $q=4\pi/5$ and amplitude $\varepsilon = 0.1$ is added to give the sinusoidal profiles in (a)--(c) at $t=0$.  Results in (d)--(f) show the solution of Equations (\ref{Eq:PFSphi})--(\ref{Eq:PFSphi MovingBC})  at $t=50$ where, in each case, we can see that each front moves in the positive $x$--direction and  amplitude of the transverse perturbation clearly decays.  Plots in (g)--(i) show the evolution of $\partial \Omega(t)$ at $t=0, 2, 4, 6, 8$ with the arrow showing the increasing value of $t$.  Note that the plots in (g)--(i) focus on the interval $4 \le x \le 6$ so that the details of the interface evolution are clear.  Comparing these plots of $\partial \Omega(t)$ suggests that these travelling waves with $\kappa = 0.1$ are stable as the perturbation amplitude appears to decay with time.  All numerical calculations are performed on a $10 \times 10$ domain uniformly discretised with a $201 \times 201$ mesh.}
\end{figure}

Profiles in Figure \ref{Fig5}(a)--(c) show three travelling wave solutions for $\kappa = 0.1$ with $m=0.5, 1.0, 2.0$, respectively, where a small transverse perturbation is added in each case.  We solve Equations (\ref{Eq:PFSphi})--(\ref{Eq:PFSphi MovingBC}) with the level--set method with the profiles in Figure \ref{Fig5}(a)--(c) as the initial condition.  These solutions evolve to the profiles shown in Figure \ref{Fig5}(d)--(f) at $t=50$ where we see that the amplitude of the perturbation decays so that the moving boundary is almost perfectly vertical.  Additional profiles in Figure \ref{Fig5}(g)--(i) show the time evolution of $\partial \Omega(t)$.  Comparing the evolution of the shape of the moving boundary indicates that the amplitude of the initial perturbation decays with time indicating that  these invading travelling wave profiles are stable when a small transverse perturbation is added.

\begin{figure}
\includegraphics[width=1.0\textwidth]{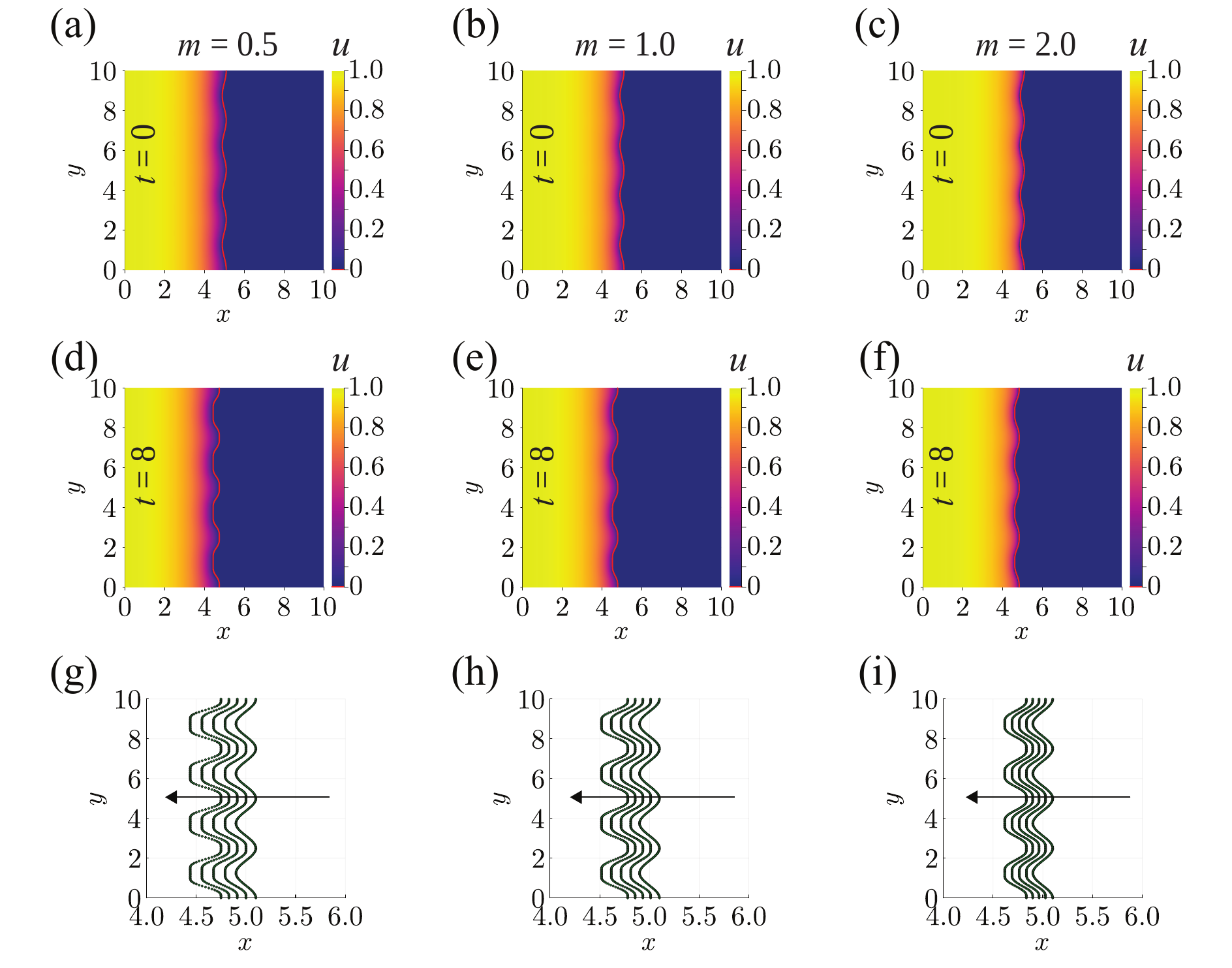}
\caption{\label{Fig6} Numerical exploration of front stability with $\kappa = -0.1$.  Profiles in (a)--(c) show three perturbed travelling wave profiles for $m=0.5, 1.0, 2.0$, respectively.  Each perturbed travelling wave profile is computed by first solving the $\mathcal{O}(1)$ boundary value problem to give the shape of the unperturbed travelling wave that is then shifted so that $\partial \Omega(0)$ coincides with the vertical line at $x=5$ before a perturbation with wavenumber $q=4\pi/5$ and amplitude $\varepsilon = 0.1$ is added to give the sinusoidal profiles in (a)--(c) at $t=0$.  Results in (d)--(f) show the solution of Equations (\ref{Eq:PFSphi})--(\ref{Eq:PFSphi MovingBC})  at $t=50$ where, in each case, we can see that each front moves in the positive $x$--direction and  amplitude of the transverse perturbation clearly decays.  Plots in (g)--(i) show the evolution of $\partial \Omega(t)$ at $t=0, 2, 4, 6, 8$ with the arrow showing the increasing value of $t$. Note that the plots in (g)--(i) focus on the interval $4 \le x \le 6$ so that the details of the interface evolution are clear.  Comparing these plots of $\partial \Omega(t)$ suggests that these travelling waves with $\kappa = -0.1$ are unstable as the perturbation amplitude appears to grow with time.  All numerical calculations are performed on a $10 \times 10$ domain uniformly discretised with a $201 \times 201$ mesh.}
\end{figure}

\newpage
Profiles in Figure \ref{Fig6}(a)--(c) shows three travelling wave solutions for $\kappa = -0.1$ with $m=0.5, 1.0, 2.0$, respectively, where a transverse perturbation is added in the same way as in Figure \ref{Fig5}(a)--(c).  Using our level--set code we solve Equations (\ref{Eq:PFSphi})--(\ref{Eq:PFSphi MovingBC}) using the profiles in Figure \ref{Fig6}(a)--(c) as the initial condition to give the profiles in Figure \ref{Fig6}(d)--(f) at $t=8$ where we see that the amplitude of the transverse perturbation does not decay with time.  Figure \ref{Fig6}(g)--(i) shows the evolution of  $\partial \Omega(t)$ where we see that the transverse perturbation does not decay, instead the shape of the wave develops into a nonlinear scallop--shape as the wavefront moves in the negative $x$--direction.  In summary, our preliminary numerical explorations in Figures \ref{Fig5}--\ref{Fig6} suggest that invading travelling waves with $\kappa > 0$ are stable, whereas receding travelling waves with $\kappa < 0$ are unstable when a small transverse perturbation is added.

\subsection{Travelling wavefront stability: Linear stability analysis} \label{Sec:Stability_LSA}
We now provide more mathematical insight into the preliminary numerical simulations in Figures \ref{Fig5}--\ref{Fig6} by carrying our a linear stability analysis following the approach of Muller and van Saarloos~\cite{Muller2002}.  To proceed we consider Equations (\ref{Eq:PFSLevelSet})--(\ref{Eq:PFSLevelSet_Extension}) in terms of the original density variable, $u$,
\begin{subequations}
 \label{level_system_u}
\begin{align}
    \frac{\partial u}{\partial t}&= \frac{\partial}{\partial x}\left[u^m\;\frac{\partial u}{\partial x}\right]+ \frac{\partial}{\partial y}\left[u^m\;\frac{\partial u}{\partial y}\right] + u(1-u)\;\;\label{PDE_levelset}\; \; \text{on}   \;\;   \phi(x,y,t)<0,\\
    &\frac{\partial \phi}{\partial t}+F\mid\nabla\phi\mid =0\;\;\text{on}   \;\;   (x,y)\in\mathcal{D},\\
    &u(x,y,t)=0 \;\; \text{on}   \;\;  \phi(x,y,t)=0,\\
    &u(x,y,0)=U(x,y) \;\; \text{on}   \;\; \phi(x,y,0)<0,\\
&\mathcal{V}=-\kappa\;u^m\;\nabla u \cdot \hat{n}\;\;  \text{on}   \;\;  \phi(x,y,t)=0. \label{PDE_levelset_boundary}
\end{align}
\end{subequations}
As we have demonstrated numerically in Figure \ref{Fig2}, the long--time solution of this system with appropriate boundary conditions leads to travelling wave solutions whose properties are consistent with our phase plane analysis presented in Section \ref{Sec:PP}.  To describe the perturbed travelling wave initial condition shown in Figures \ref{Fig5}--\ref{Fig6} we express the horizontal position of the travelling wave front, denoted $x=L(y,t)$, in terms of the travelling wave speed and a small transverse perturbation,
\begin{equation}
\label{eq:results_perturbation_front}
L(y,t) = ct + \varepsilon \exp({\mathrm{i} qy + \omega t}) + \mathcal{O}(\varepsilon^2),
\end{equation}
were $q$  is the wave number and $\varepsilon \ll 1$ is the amplitude of the small perturbation~\cite{Muller2002,Tam2023}.  Here $\omega$ is an unknown constant that we will use to describe the short time decay $\omega < 0$ or growth $\omega > 0$ of the transverse perturbation amplitude.  Our aim now is to assume that we have $\varepsilon \ll 1$ and use this small parameter to derive a perturbation solution to explore the dispersion relationship $\omega(q)$ for a range of travelling wave solutions characterised by different choices of $\kappa$ and $m$ which, as we showed in Section \ref{Sec:PP} implicitly defines the associated travelling wave speed $c$.  This approach involves numerically solving a series of boundary value problems with one independent variable for the real component of the growth rate $\omega$.

To proceed, we write
\begin{equation}
\label{eq:variable_z}
\xi = x - ct - \varepsilon\exp(\mathrm{i} qy + \omega t) + \mathcal{O}(\varepsilon^2),
\end{equation}
and we also perturb the signed distance function $\varphi = x - ct - \varepsilon\exp(\mathrm{i} qy + \omega t) + \mathcal{O}(\varepsilon^2)$ to be consistent with our perturbed density function.  Expanding $u(x,y,t)$ gives
\begin{equation}
\label{eq:perturbation_density_u_v_u0}
u(x,y,t) = u_0(\xi) + \varepsilon u_1(\xi)\exp(\mathrm{i} qy + \omega t) + \mathcal{O}(\varepsilon^2).
\end{equation}
Substituting these expansions into the level-set formulation gives, to leading order, the same boundary value problem for the travelling wave solution that we considered in Section \ref{Sec:PP}, namely
\begin{subequations}
    \label{level_system_xi_zero}
\begin{align}
    -\left(\frac{\mathrm{d} u_0}{\mathrm{d}\xi}\right) c&=\left(\frac{\mathrm{d}^{2}u_0}{\mathrm{d}\xi^{2}}\right) u_0^m+m\; \left(\frac{\mathrm{d}u_0}{\mathrm{d}\xi}\right)^{2} u_0^{m-1} - u_0\left(u_0-1\right) \;\; \text{on}   \;\;   -\infty < \xi<0, \label{eq:LeadingorderBVP}\\
    &\lim_{\xi \to -\infty} u_0(\xi)=1,\\
    &u_0=0 \; \text{at}   \;\;   \xi=0,\\
&-c=\kappa u_0^m \left(\frac{\mathrm{d}u_0}{\mathrm{d}\xi}\right) \; \text{at}   \;\;   \xi=0. \label{eq:LeadingorderBVPBC}
\end{align}
\end{subequations}
The solution of this boundary value problem gives $u_0(\xi)$, which describes the shape of the long--time travelling wave solution. Full details of a shooting method to solve this boundary value problem for $u_0(\xi)$, $c$ and $\kappa$ are provided in the Appendix.

Given that the solution of the leading order problem for $u_0(\xi)$, $c$ and $\kappa$, we then consider the $\mathcal{O}(\varepsilon)$ boundary value problem for $u_1(\xi)$,
\begin{subequations}
\begin{align}
    -\omega \left(\frac{\mathrm{d} u_0}{\mathrm{d}\xi}\right)+\omega u_1-\left(\frac{\mathrm{d} u_1}{\mathrm{d}\xi}\right) c&= u_0^{m-1}m\left(\frac{\mathrm{d}^2u_0}{\mathrm{d}\xi^2}\right) u_1
+ \left(\frac{\mathrm{d}^2u_1}{\mathrm{d}\xi^2}\right) u_0^m+m\left(\frac{\mathrm{d} u_0}{\mathrm{d}\xi}\right)^2u_1(m-1)u_0^{m-2}\nonumber\\
&+ \left(q^{2} u_0^m+2 u_0^{m-1}m \frac{{\mathrm{d}}u_1}{d\xi}\right) \left(\frac{\mathrm{d} u_0}{\mathrm{d}\xi}\right)
-u_1 \left(q^2u_0^m-2u_0-1\right) \;\; \text{on}   \;\;   -\infty < \xi<0, \label{eq:OrdereBVP} \\
    &\lim_{\xi \to -\infty} u_1(\xi)=0,\\
    &u_1=0 \;\; \text{at}   \;\;   \xi=0, \\
&-\omega=\kappa u_0^{m-1}\left(u_0 \left(\frac{\mathrm{d}u_1}{\mathrm{d}\xi}\right)+u_1 \left(\frac{\mathrm{d}u_0}{\mathrm{d}\xi}\right)\right)\;\;  \text{at}   \;\;  \xi=0. \label{eq:OrdereBVP2}
\end{align}
\end{subequations}
To solve the $\mathcal{O}(\varepsilon)$ boundary value problem for $u_1(\xi)$ and $\omega$ we must specify the perturbation wavenumber $q$.  Repeatedly solving the $\mathcal{O}(\varepsilon)$  problem for different values of $q$ allows us to construct the dispersion relationship, $\omega(q)$ from which we can determine the linear stability of the perturbation.  Details of a shooting method to solve the $\mathcal{O}(\varepsilon)$ boundary value problem are given in the Appendix.

With these tools we can now calculate the dispersion relationship $\omega(q)$ for various travelling wave solutions where the front is perturbed by a small sinusoidal transverse perturbation with wavenumber $q$.  Linear stability results can be compared with full 2D time--dependent PDE solutions obtained using the level--set method outlined in Section \ref{Sec:NumericalMethod}.  To use the 2D time--dependent PDE model as a check on the accuracy of our linear stability analysis we use the solution of the boundary value problems for $u_0(\xi)$ and $u_1(\xi)$ to define appropriate initial conditions for the level--set method,
\begin{subequations} \label{eq:perturbedic}
\begin{align}
u(x, y, 0) &= u_0\left(x - \beta - \varepsilon\cos(qy)\right) + \varepsilon u_1\left(x - \beta - \varepsilon\cos(qy)\right)\cos(qy),\label{eq:results_numerical_ic_u}\\
\phi(x, y, 0) &= x - \beta - \varepsilon\cos(qy),\label{eq:results_numerical_ic_phi}
\end{align}
\end{subequations}
where $\beta$ is a constant that we can choose so that the perturbed travelling wave front it initially positioned half way along the horizontal extent of the $\mathcal{D}$ as we did in Figures \ref{Fig5}--\ref{Fig6}.  Solving Equations (\ref{Eq:PFS})--(\ref{Eq:PFS MovingBC}) with these initial conditions and the corresponding value of $\kappa$ and $m$ gives us the full time evolution of $u(x,y,t)$ from which we can estimate the short--time growth or decay of the perturbation amplitude~\cite{Muller2002,Tam2023}.  At each time point that we record $u(x,y,t)$, we estimate the perturbation amplitude as follows,
\begin{subequations}
\begin{align}
A(t) &= \frac{X_{\textrm{max}}(t) - X_{\textrm{min}}(t)}{2}, \text{where},\\
X_{\textrm{max}}(t) &= \max_{y}[x \mid \phi(x, y, t) = 0], \\
X_{\textrm{min}}(t) &= \min_{y}[x \mid \phi(x, y, t) = 0].
\end{align}
\end{subequations}
To compute these quantities we use linear interpolation to find positions $x_j$ such that $\phi(x, y_j, t_k) = 0$ for the $j$th row in the discretisation of $\mathcal{D}$ at time $t = t_k$. The maximum and minimum values in the set ${x_j}$ determine $X_{\textrm{max}}(t)$ and $X_{\textrm{min}}(t)$, respectively.
To estimate $\omega$ we use Equation (\ref{eq:results_perturbation_front}) and assume that the perturbation amplitude grows or decays exponentially fast over a sufficiently short time interval.  We use the Polynomials.jl package in Julia to perform a least--squares regression $\log(A) = \omega_{\textrm{n}} t + C$, where $C$ is a constant, and $\omega_{\textrm{n}}$ is the numerical estimate of the growth rate.  For all results presented we consider a relatively short interval $t \in [1,2]$ with 21 equally--spaced time points within that interval to compute $\omega_{\textrm{n}}$.  For each problem considered we generate a plot of $\log(A)$ as a function of $t$ which allows us to check that $A(t)$ grows or decays exponentially with time, as indicated by our data falling on a straight line over the time interval considered.

\begin{figure}
\includegraphics[width=1.0\textwidth]{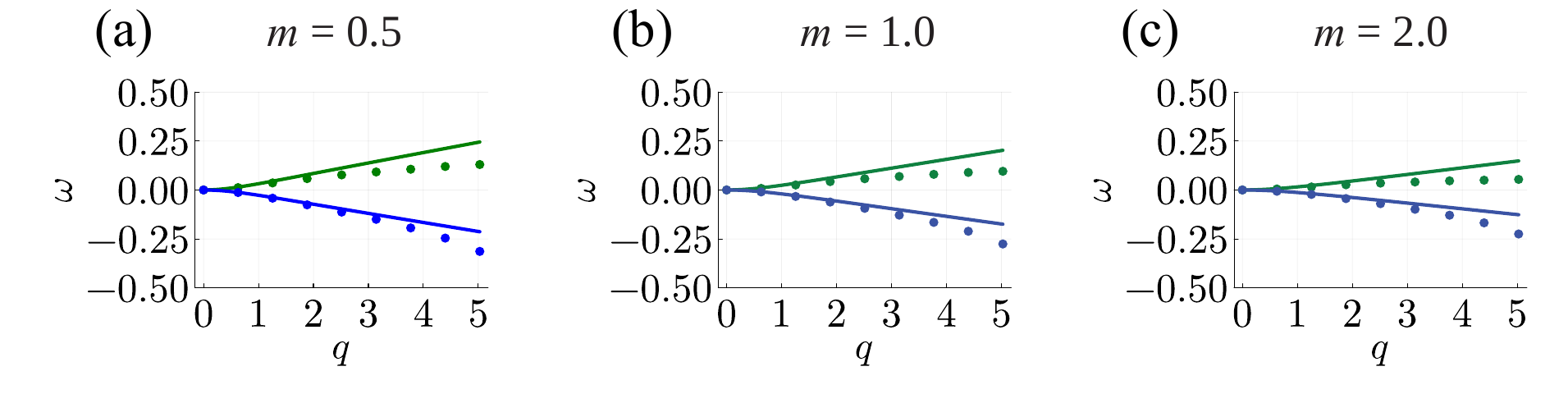}
\caption{\label{Fig7} Dispersion relationships showing $\omega(q)$ for $m=0.5, 1.0, 2.0$, as indicated.  Each plot shows the dispersion relationship derived using linear stability analysis for invading travelling waves with $\kappa = 0.1$ (solid blue lines) and receding travelling waves with $\kappa = -0.1$ (solid green lines), indicating that invading travelling waves with $\kappa > 0$ and $c > 0$ are linearly stable, whereas receding travelling waves with $\kappa < 0$ and $c<0$ are linearly unstable.  The linear stability results are compared with estimates of the growth rate from the full 2D time--dependent PDE solutions (discrete discs) computed on a domain with $0 \le x \le 10$ and $0 \le y \le 10$ domain that is uniformly discretised using a $201\times201$ mesh.}
\end{figure}

\begin{figure}
\includegraphics[width=1.0\textwidth]{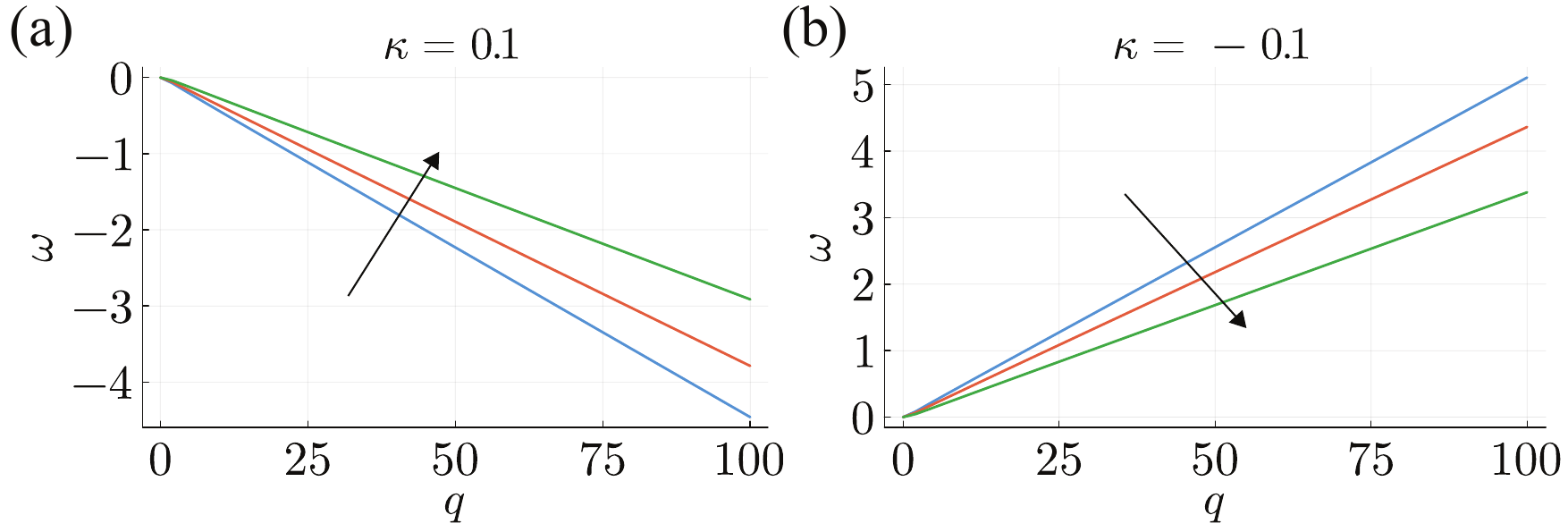}
\caption{\label{Fig8}   Dispersion relationship $\omega(q)$ for larger wavenumbers using linear stability analysis.  Results in (a)--(b) correspond to $\kappa =0.1$ and $\kappa = -0.1$, respectively.  Each plot shows the dispersion relationship for $m=0.5, 1.0, 2.0$ in blue, red and green, respectively, and the arrow shows the direction of increasing $m$.}
\end{figure}

Results in Figure \ref{Fig7} compare the dispersion relationship $\omega(q)$ from the linear stability analysis with estimates of the growth rates using the full 2D time--dependent PDE solution.  In general we see that the linear stability analysis is consistent with the preliminary results in Figures \ref{Fig5}--\ref{Fig6} since invading travelling waves ($\kappa > 0$, $c > 0$) are linearly stable to small transverse perturbations whereas receding travelling waves ($\kappa < 0$, $c < 0$) are linearly unstable to transverse perturbations.  This is the first time that the stability of travelling wave solutions to the PFS model have been considered. Our results indicate that the agreement between our estimates of $\omega(q)$ using the two approaches is quite good, especially for small wavenumbers, $q < 2$.  Interestingly, the match between $\omega(q)$ from the linear stability analysis and the time--dependent PDE solutions becomes increasingly poorer for larger $q$.  This trend has a simple but important explanation since all time--dependent PDE results in Figure \ref{Fig7} are obtained using the same spatial discretisation of $\mathcal{D}$ regardless of $q$.  We expect that the 2D mesh would need to be refined to capture the increasingly fine details of the perturbed initial condition $u(x,y,0)$ and the subsequent evolution of $u(x,y,t)$ as $q$ increases.   From this point of view, the linear stability analysis is computationally advantageous since this approach requires us to solve two boundary value problems with just one independent variable, $-\infty < \xi < 0$ which is straightforward to discretise.  Furthermore, the discretisation of $-\infty < \xi < 0$ is completely independent of $q$.   Therefore, should we wish to explore the dispersion relationship for larger $q$ it is preferable to use the linear stability approach.  Additional results in Figure \ref{Fig8} show that the dispersion relationship computed for much larger wavenumbers using the linear stability approach.  Here we see that $\omega(q)$ decreases approximately linearly with $q$ for invading travelling waves with $\kappa > 0$ and $c > 0$, whereas $\omega(q)$ increases approximately linearly with $q$, for receding travelling waves with $\kappa < 0$ and $c < 0$.   These extended results in Figure \ref{Fig8} for larger $q$ indicate the approximately linear $\omega(q)$ plots for small $q$ in Figure \ref{Fig7} also hold for larger wavenumbers which are computationally infeasible to explore using the full time--dependent PDE model.

Results in Figure \ref{Fig6}--\ref{Fig7} allow us to briefly return to Section \ref{Sec:Intro} where, in Figure \ref{Fig1}(e)--(h) where we hypothesized four potential outcomes in terms of invading, receding, linearly stable and linearly unstable travelling waves.  Our results in Figure \ref{Fig7}--\ref{Fig8} indicate that invading travelling waves with $\kappa = 0.1$ and $m=1$ are linearly stable while receding travelling waves with $\kappa =-0.1$ and $m=1$ are linearly unstable.  We have found that these broad observations hold for other choices of $m$ and $\kappa$, namely that setting $\kappa > 0$ leads to invading travelling waves that are linearly stable and setting $\kappa < 0$ leads to receding travelling waves that are linearly unstable.  Our open-source Julia codes provided on \href{https://github.com/alex-tam/OnePhase_PorousFisherStefanStability}{Github} can be used to explore these other parameter choices.  In summary, of the four potential outcomes previously hypothesized in Figure \ref{Fig1}(e)--(h) we have only found evidence to support the scenario depicted in Figure \ref{Fig1}(e) for $\kappa > 0$ and Figure \ref{Fig1}(h) for $\kappa < 0$.

\section{Conclusion and future work}
In this work we re--visited the recently--established Porous--Fisher--Stefan (PFS) model of biological invasion, and focused our attention on properties of the travelling wave solutions of that model.  The first presentation of the PFS model focused on working with a degenerate nonlinear diffusion term with flux $\mathcal{J} = -u \partial u / \partial x$ on $-\infty < x < L(t)$ with $\textrm{d}L(t)/\textrm{d} t = -\kappa u \partial u/\partial x$ at $x=L(t)$, and the existence of travelling waves describing biological invasion with $\kappa > 0$ was established using phase plane and perturbation methods~\cite{Fadai2020}.  Here we extend this modelling framework by dealing with a 2D model that incorporates a generalised degenerate diffusion term with flux $-u^m \nabla u$, and we consider both invading fronts with $\kappa >0$, and receding fronts with $\kappa < 0$. Solving the 2D model in an appropriate domain provides strong numerical evidence that the model supports long--time travelling wave solutions, and working in the phase plane allows us to examine properties of the travelling wave solutions over a range of model parameters in the PDE model, and to determine how the long--time travelling wave speed $c$ depends upon the choice of parameters in the PDE model.

Preliminary explorations of the time--dependent PDE solutions indicate that invading travelling waves with $\kappa > 0$ and $c > 0$ are stable when a small transverse perturbation with wavenumber $q$ is introduced at the moving boundary, whereas receding travelling waves with $\kappa < 0$ and $c< 0$ are unstable.  These informal numerical observations are corroborated with a linear stability analysis indicating that the short--time amplitude of the transverse perturbation increases when $\kappa < 0$, whereas it decreases with $\kappa > 0$.  In general we find reasonably good agreement between the dispersion relationship obtained from the linear stability analysis and from the 2D level--set method, especially for small $q$.  For larger $q$ it is straightforward to carry out the linear stability analysis because this involves solving two boundary value problems with just one independent variable that we discretise efficiently using a non-uniform mesh.  In contrast, solving the 2D time--dependent PDE problem for larger $q$ requires an increasingly refined spatial mesh as $q$ increases, which makes the time--dependent PDE solutions impractical.

As illustrated in Figures \ref{Fig7}--\ref{Fig8} setting $\kappa < 0$ leads to linearly unstable travelling wave solutions.  In the heat transfer literature, linearly unstable front propagation is often studied using regularisation by including additional mechanisms in the PDE model that can stabilise perturbation modes with large wavenumbers~\cite{Chadam1983,Yang2002}.  A standard approach in heat and mass transfer for Stefan--type moving boundary problems is to include terms describing surface tension effects~\cite{Chadam1983}.  Introducing surface tension involves replacing the usual 1D interface condition $u(L(t),t)=0$ with $u(L(t),t)=\gamma \mathcal{K}$, where $\gamma > 0$ is the surface tension coefficient, and $\mathcal{K}$ is the signed curvature of the moving front~\cite{Chadam1983}.    This approach is reasonable in heat transfer and fluid mechanics where the dependent variable represents temperature and pressure, respectively.  This means that the boundary condition $u(L(t),t)=0$  corresponds to a zero scaled temperature or zero scaled reference pressure, and it is physically reasonable to consider a negative scaled temperature or pressure at the 2D moving front with negative curvature.  Working with the PFS model is very difference since here $u$ represents a scaled population density which means that it is physically unreasonable to set $u(L(t),t)=\gamma \mathcal{K}$ which would be negative at certain points on the 2D moving front with negative curvature. Therefore, in this work we have not pursued any regularisation of the linearly unstable travelling wave fronts as the classical approach is not suitable for this application.

There are many opportunities for future work motivated by the problems that we have considered here.   For example, one question for future consideration is to explore different options for biologically meaningful regularisation terms that could be applied both to linearly unstable front propagation for the PFS model, as well as other moving boundary problems that describe population biology phenomena.  Other technical challenges could also be considered.  For example, our phase plane analysis in Section \ref{Sec:PP} led to expressions for an exact solution for $c=0$ and $m=1$, which we used to construct perturbation solutions for $|c| \ll 1$.  These solutions are insightful because they can be used to provide insight into the relationship between $\kappa$ and $c$ for $m=1$.  Another option would be to use these solutions to construct perturbation solutions for $|m-1| \ll 1$ which could be used to provide insight into the relationship between $\kappa$ and $c$ for $m \ne 1$.  We leave both of these extensions for future consideration.

\section*{Appendix: Numerical solutions of various boundary value problems}\label{Numerical_BVP}
\subsection*{Numerical solution of the $\mathcal{O}(1)$ boundary value problem}
We solve Equations (\ref{eq:LeadingorderBVP})--(\ref{eq:LeadingorderBVPBC}) by re--writing the boundary value problem in terms of $\phi_0 = u_0^{(1+m)}$,
\begin{subequations}
\begin{align}
    &\phi_0^{m/(1+m)} \left(\frac{\mathrm{d}^{2}\phi_0}{\mathrm{d}\xi^{2}}\right) +c\left(\frac{\mathrm{d} \phi_0}{\mathrm{d}\xi}\right) +(1+m) \phi_0\left(1-\phi_0^{1/(1+m)}\right)=0\; \; \text{on}   \;\; -\infty <  \xi<0,\\
    &\lim_{\xi \to -\infty} \phi_0(\xi)=1,\\
    &\phi_0=0\;\;  \text{at}   \;\;  \xi=0,\\
\frac{\mathrm{d}\phi_0}{\mathrm{d}\xi}&=-\frac{c(1+m)}{\kappa}  \;\;  \text{at}   \;\;  \xi=0. \label{eq:NondimbcDirichlet_2D_phi_m=1_level_uu}
\end{align}
\end{subequations}
To proceed we approximate the infinite domain by some sufficiently large truncated finite  domain,
\begin{subequations}
    \begin{align}
    &\phi_0^{m/(1+m)} \left(\frac{\mathrm{d}^{2}\phi_0}{\mathrm{d}\xi^{2}}\right) +c\left(\frac{\mathrm{d} \phi_0}{\mathrm{d}\xi}\right) +(1+m) \phi_0\left(1-\phi_0^{1/(1+m)}\right)=0\;\; \text{on}   \;\; -\xi_{\textrm{max}}<  \xi<0, \label{eq:LeadingOrder1}\\
    & \phi_0=1 \;\; \text{at}   \;\;  \xi=-\xi_{\textrm{max}} \\
    &\phi_0=0 \;\;  \text{at}   \;\;  \xi=0,   \label{eq:LeadingOrder2}\\
    \frac{\mathrm{d}\phi_0}{\mathrm{d}\xi}&=-\frac{c(1+m)}{\kappa}  \;\;  \text{at}   \;\;  \xi=0,
\end{align}
\end{subequations}
where the unknown travelling wave speed $c$ will be determined using a shooting procedure that we now describe.

In general, $\phi_0$ varies rapidly with $\xi$ near $\xi=0$, whereas $\phi_0$ approaches unity as $\xi \to -\xi_{\textrm{max}}$.  We take advantage of these properties by employing a non-uniform mesh where the mesh resolution is fine near $\xi=0$ and becomes increasingly coarse near $\xi = -\xi_{\textrm{max}}$.  One way of achieving this is to use a geometrically spaced mesh~\cite{McCue2022} where the mesh spacing varies from $\delta\xi_{\textrm{min}}$ at $\xi = 0$ to $\delta \xi_{\textrm{max}}$ at $\xi=-\xi_{\textrm{max}} = -20$ using $N$ geometrically--spaced  mesh points.  At the $i$th internal mesh points we have $\delta\xi_{i} = (1+\epsilon) \delta\xi_{i+1}$.  For $\delta\xi_{\textrm{min}} = 1 \times 10^{-5}$ with $N=301$ mesh points we obtain a geometric expansion factor of $(1+\epsilon) = 1.0381894$ and  $\delta\xi_{\textrm{max}}=0.7357$.    Numerical experimentation confirms that this nonuniform mesh spacing leads to grid--independent results for the problems that we consider.

At the $i$th internal mesh point we define $h^{+}_i=\xi_{i+1}-\xi_{i}$ and  $h^{-}_i=\xi_{i}-\xi_{i-1}$ and discretise Equations(\ref{eq:LeadingOrder1})--(\ref{eq:LeadingOrder2}) on the non-uniform mesh leading to a system of $N$ nonlinear algebraic equations $\vect{F}(\phi_{0}^{(i)}) = \vect{0}$ which we solve using Newton-Raphson iteration.  Elements of the vector $\vect{F}$ are given by,
\begin{subequations}
    \begin{align}
        F_0 &= \phi_{0}^{(0)} - 1,\\
        F_i &={\phi_0^{(i)}}^{m/(1+m)} \left(A \phi_0^{(i+1)} +B \phi_0^{(i)} + C\phi_0^{(i-1)}\right) + c\frac{\phi_0^{(i+1)} - \phi_0^{(i-1)}}{h_i^++h_i^-} +(1+m) \phi_0^{(i)}\left(1-{\phi_0^{(i)}}^{1/(1+m)}\right), \\
        F_{N} &= \phi_0^{(N)}\;,
    \end{align}
\end{subequations}
where the superscript $i = 2, \dots, N-1$  refers to the $i$th internal mesh point, and the coefficients $A= {2}/({h^+_i(h_i^++h_i^-)}) $, $B={-2}/({h_i^+h_i^-}) $, and $C= {2}/({h^-_i(h_i^++h_i^-)})$ arise by applying standard finite difference stencils on the nonuniform mesh~\cite{McCue2022}.  An initial estimate of $\vect{\phi}_0^{(0)}$ can be updated using Newton-Raphson iteration as follows,
\begin{equation}
    \label{eq:newton_method_u0_non_uniform}%
    \vect{\phi}_0^{(n+1)} = \vect{\phi}_0^{(n)} - J^{-1}\left(\vect{\phi}_0^{(n)}\right)\vect{F}\left(\vect{\phi}_0^{(n)}\right),
\end{equation}
where the superscript $n$ gives the iteration count and $J_{ij} = \partial_j F_i$ are the elements of the Jacobian matrix for $\vect{F}(\phi_{0}^{(i)})$.   Iterations are terminated when $\lVert \vect{\phi}_0^{(n+1)} - \vect{\phi}_0^{(n)} \rVert < 1 \times 10^{-6}$.  In this work we take $\vect{\phi}_0^{(0)}$ to be a piecewise linear approximation of the shape of the travelling wave profile.

To estimate the travelling wave speed $c$ we use a shooting method that commences with an initial estimate of $c^{(0)}$  to solve the leading order boundary value problem for $\phi_0(\xi)$. For our estimates of $c^{(0)}$ and $\phi_0(\xi)$ we discretise the moving boundary condition, Equation (\ref{eq:NondimbcDirichlet_2D_phi_m=1_level_uu}), on our non-uniform mesh to give
\begin{equation}
    \label{eq:discretised_Stefan_condition}%
    f^{(k)} = c^{(k)} + \frac{\kappa}{m+1} \left(P \phi_{0}^{({N-2},k)}-Q \phi_{0}^{({N-1},k)}+R \phi_{0}^{({N},k)}\right),
\end{equation}
where $P=h^+_{N-1}/({h^-_{N-1}(h^+_{N-1}+h^-_{N-1})})$, $Q= (h^+_{N-1}+h^-_{N-1})/(h^+_{N-1} h^-_{N-1})$,, and  $R= (2h^+_{N-1}+h^-_{N-1})/({h^+_{N-1}(h^+_{N-1}+h^-_{N-1})})$ is the estimate of the wave speed after $k$ iterations and while $\phi_{0}^{(i, k)}$ denotes our estimate of $\phi_0$ at the $i$th mesh point after $k$ iterations of the shooting method. To satisfy Equation (\ref{eq:NondimbcDirichlet_2D_phi_m=1_level_uu}) we update our estimate of $c$ so that $f = 0$, again using Newton-Raphson iteration,
\begin{equation}
    \label{eq:newton_c}%
    c^{(k+1)} = c^{(k)} - \frac{f\left(c^{(k)}\right)}{f'\left(c^{(k)}\right)},
\end{equation}
until $\lVert c^{(k+1)} - c^{(k)} \rVert < 1 \times 10^{-6}$. To implement this algorithm we estimate $f'$ using a finite difference approximation,
\begin{equation}
    \label{eq:f_approx}
    \frac{\textrm{d}f}{\textrm{d}c} = \frac{f(c + \delta c) - f(c)}{\delta c},
\end{equation}
with $\delta c = 1 \times 10^{-6}$.  This method for computing the solution of the $\mathcal{O}(1)$ boundary value problem requires that we specify $\kappa$ and then estimate $c$ iteratively by shooting.  The efficiency of this approach strongly depends upon providing a good initial estimate of $c$ for our choice of $\kappa$ and $m$, and there are many ways to make a good estimate.  One way is to use the phase plane tools described in Section \ref{Sec:PP} to calculate an accurate estimate of $c$ for a particular choice of $m$ and $\kappa$.  A much simpler option is to use Equation (\ref{eq:c_kappa_perturbed_relation}) to estimate $c$ for a given value of $\kappa$.  Note that  Equation (\ref{eq:c_kappa_perturbed_relation}) is exact for $m=1$ and $|c| \ll 1$ only, however this relationship still provides a reasonable initial estimate of $c$ that is then iteratively improved by shooting.

\subsection*{Numerical solution of the $\mathcal{O}(\varepsilon)$ boundary value problems}
To solve for the  $\mathcal{O}(\varepsilon)$ term in the linear stability analysis we re--write Equations (\ref{eq:OrdereBVP})--(\ref{eq:OrdereBVP2}) in terms of $\phi_1 = u_0^{m}u_1$, and then consider the resulting boundary value problem on an appropriately truncated domain, which gives,
\begin{subequations}
\begin{align}
   &\left(\frac{\mathrm{d}^2\phi_1}{\mathrm{d}\xi^2}\right) u_0^{m+1} -c\phi_1\;m\left(\frac{\mathrm{d} u_0}{\mathrm{d}\xi}\right)-\omega \phi_1 u_0+cu_0\left(\frac{\mathrm{d} \phi_1}{\mathrm{d}\xi}\right)\notag \\
   &- \phi_1 \left( q^{2}u_0^{m+1}+2 u_0^2-u_0\right) = - u_0^{m+1}  \left(\frac{\mathrm{d} u_0}{\mathrm{d}\xi}\right)\left(\omega+q^2u_0^m\right) \;\; \text{on}     \;\;   -\xi_{\textrm{max}}<  \xi<0, \label{eq:BVPe1}\\
    &\phi_1=0  \;\;  \text{at}   \;\;  \xi = -\xi_{\textrm{max}},\label{eq:BVPe2}\\
     &\phi_1=0\;\;  \text{at}   \;\;  \xi=0,\label{eq:BVPe3}\\
\frac{\mathrm{d}\phi_1}{\mathrm{d}\xi}&=-\frac{\omega}{\kappa}  \;\;  \text{at}   \;\;  \xi=0. \label{eq:BVPe4}
\end{align}
\end{subequations}
Our strategy for solving the  $\mathcal{O}(\varepsilon)$ boundary value problem follows closely from our approach for solving the  $\mathcal{O}(1)$ boundary value problem by discretizing Equations (\ref{eq:BVPe1})--(\ref{eq:BVPe3}) on the same non-uniform mesh used previously to give
\begin{subequations}
  \begin{gather}
    \phi_1^{(0)} = 0,\label{eq:BVPe1d}\\
    \begin{gathered}
     \left(u_0^{(i)}\right)^{(m+1)} \left(A \phi_1^{(i+1)} +B \phi_1^{(i)} + C \phi_1^{(i-1)}\right) + cu_0^{(i)}\dfrac{\phi_1^{(i+1)} - \phi_1^{(i-1)}}{h_i^+ +h_i^-}  -c\theta_1^{(i)}\dfrac{u_0^{(i+1)} - u_0^{(i-1)}}{h_i^+ + h_i^-}\\
-\omega\;\phi_1^{(i)}\;u_0^{(i)}  -\phi_1^{(i)} \left( {(u_0^{(i)})}^{(m+1)}q^{2}+2 {(u_0^{(i)})}^2-u_0^{(i)}\right)= -\left(u_0^{(i)}\right)^{(m+1)}  \dfrac{u_0^{(i+1)} - u_0^{(i-1)}}{h_i^+ + h_i^-} \left(\omega+q^2{u^{(i)}_0}^m\right),
    \end{gathered}\label{eq:BVPe2d}\\
    \phi_1^{(N)} = 0, \label{eq:BVPe3d}
  \end{gather}
\end{subequations}
where the superscript $i = 2, \dots, N-1$ refers to the $i$th internal mesh point, and the coefficients $A= {2}/({h^+_i(h_i^++h_i^-)}) $, $B={-2}/({h_i^+h_i^-}) $. and $C= {2}/({h^-_i(h_i^++h_i^-)})$ arise by applying standard finite difference stencils on the nonuniform mesh~\cite{McCue2022}.  Values of $c$ and $u_0$ are known inputs that we have previously calculated as a result of solving the $\mathcal{O}(1)$ boundary value problem.  Similarly,  the wavenumber $q$ is also a specified input.  Here we note that Equations (\ref{eq:BVPe1d})--(\ref{eq:BVPe3d}) are linear in $\phi_1^{(i)}$ so we solve this linear system using Julia's backslash operator.

To compute $\omega$ we use a shooting method with an initial estimate of $\omega^{(0)}$ and then discretise Equation (\ref{eq:BVPe4}) on the same nonuniform mesh to give
\begin{equation} \label{eq:omega_condition}%
g^{(k)} = \omega^{(k)} + \kappa\left(P \phi_{1}^{({N-2},k)}-Q \phi_{1}^{({N-1},k)}+R \phi_{1}^{({N},k)}\right),
\end{equation}
where $P=h^+_{N-1}/({h^-_{N-1}(h^+_{N-1}+h^-_{N-1})})$, $Q= (h^+_{N-1}+h^-_{N-1})/(h^+_{N-1} h^-_{N-1})$, and  $R= (2h^+_{N-1}+h^-_{N-1})/({h^+_{N-1}(h^+_{N-1}+h^-_{N-1})})$.   To satisfy Equation (\ref{eq:BVPe4}) we update our estimate of the growth rate so that $g=0$ using Newton-Raphson iteration
\begin{equation}
    \label{eq:newton_omega}%
    \omega^{(k+1)} = \omega^{(k)} - \frac{g\left(\omega^{(k)}\right)}{g'\left(\omega^{(k)}\right)},
\end{equation}
$\lVert \omega^{(k+1)} - \omega^{(k)} \rVert < 1 \times 10^{-6}$.  To implement this algorithm we estimate $g'$ using a finite difference approximation,
\begin{equation}
    \label{eq:f_approx_g}
    \frac{\textrm{d}g}{\textrm{d}\omega} = \frac{g(\omega + \delta \omega) - g(\omega)}{\delta \omega},
\end{equation}
with $\delta \omega = 1 \times 10^{-6}$.

\end{document}